\documentclass[aps,preprint,showpacs]{revtex4}
\usepackage{graphicx}

\begin{document}
\title{A simple method to make the Wang-Landau sampling converge}
\author{Shijun Lei}
\affiliation{Purple Mountain Observatory, 2 West Beijing Rd, Nanjing Jiangsu, 210008 P.R. China}
\date{\today}

\begin{abstract}
We show that a histogram maintained throughout the Wang-Landau (WL) sampling for the energy entries visited
during the simulation could be used to make the simulated density of states (DOS) converge. The method is easy
to be implemented to the WL sampling with no extra computational cost and bears the advantages of both the WL
method and the multicanonical method.
\end{abstract}
\maketitle

Among the various simulation methods to obtain the the density of states (DOS), the WL sampling \cite{Wang01}
has been proven both efficient and robust, triggering extensive practical applications \cite{Li07,Wust,Wust12},
theoretical probing \cite{Zhou03,Zhou08} and further improvements \cite{Cunha-Netto08,Belardinelli07,Vogel13}.
However, the saturation of the error in the simulated DOS in WL sampling was also noted \cite{Yan03, Shell03}. 

A successful solution to the non convergence of the traditional WLS seems to be the $1/t$ WLS proposed by
\cite{Belardinelli07}, who showed that the error in the improved algorithm vanishes following the $1/\sqrt{t}$ law.

We provide in this article an alternative solution to the non convergence of the WL sampling. The method simply makes
use of a histogram $H_{t}(E)$ maintained throughout the simulation. Same as the histogram $H_{f}(E)$ used in WL
sampling for the updating of the modification factor $f$ when the histogram is flat enough, $H_{t}(E)$ is initialized to
be $H_{t}(E)=0, \forall E$ and is updated as $H_{t}(E_i)\leftarrow H_{t}(E_i)+1$ for the energy entry $E_i$ visited after a
simulation move. The only difference is that $H_{t}(E)$ is not refreshed to zero every time $f$ is modified, rather, it is
maintained throughout the simulation. And at the end of the simulation, we simply update the DOS as
$g(E)\leftarrow g(E)\times H_{t}(E), \forall E$. 

While the new method is extremely easy to be implemented to the WL sampling with almost no extra computational cost,
we show in the following that it solve the non convergence of the WL method and works as well as the simple MC method
in long run.

Test for the classical $8\times8$ Ising lattice, we show in Fig. 1 and 2 the behavior of the errors in the DOS and the critical
temperature ($T_C$) as a function of simulation time $t$ for various methods and the new method is labeled as "MC WL"
(see below for the naming). Similar to \cite{Wang01}, the errors in the DOS is calculated as

\begin{equation}
\epsilon_S=\frac{1}{N}\displaystyle\sum_{E}\left |\frac{S_{sim}(E)-S_{exc}(E)}{S_{exc}(E)}\right |
\label{equ1}
\end{equation}

and that in the critical temperature is

\begin{equation}
\epsilon_{T_C}=\left |\frac{T^{sim}_C-T^{exc}_C}{T^{exc}_C}\right |
\label{equ2}
\end{equation}

We start the MC simulation with an exact DOS and all the other WL like ones with $S(E)=0, \forall E$. The modification factor
$f$ is reduced according to an $80\%$ flatness criterion of the histogram $H_{t}(E)$, i.e., when every energy entry has a count
no less than $80\%$ of the average. The simulation time is defined as $t=N/63$, where $N$ is the number of simulation moves
and $63$ the total energy entries for the  $8\times8$ Ising model.

Judged by the DOS and $T_C$, it is clear that both the $1/t$ WL and the new method (MC WL) produce simulated results that
converge to the exact value and perform as well as the simple MC method in long run.

In Fig. 3 we show the the best-fit to the histograms of $T_c$ obtained from 50000 independent runs of simulations using MC,
$1/t$ WL and MC WL method. The fitting results are also summarized in the table below.The three methods are of similar accuracy
with $1/t$ WL and MC WL showing some systematic errors in $T_C$ for a given simulation time.

\begin{center}
\begin{tabular}{| l | c | r |}
\hline
Method & $T_C$ & $\sigma(T_C)$ \\ \hline
MC & 2.36190 & $6.53\times10^-3$  \\ \hline
$1/t$ WL & 2.36141 & $6.58\times10^-3$\\ \hline
MC WL & 2.36254 & $6.62\times10^-3$\\ \hline
\end{tabular}
\end{center}

Here test for a simple Ising model only, we believe the convergency of the new method is general. Actually, given the saturated
error in the WL method and the suggestion of $1/t$ scheme of reducing $f$ in long run, many people believe that to reduce the
modification factor in a manner of $f\leftarrow \sqrt{f}$ ($\ln f \leftarrow 1/2\ln f$) is too fast in long run. When $f$ turns out to be
too small to be affective in long run, the WL sampling eventually evolves into a simple MC one, where the DOS is update likewise
using the histogram. And we thus name the new method multicanonical Wang-Landau (MC WL) method. Further studies are under
going to see if the new method really bears the advantages of both methods.

\begin{figure}
\centering{\includegraphics[width=1.0\textwidth]{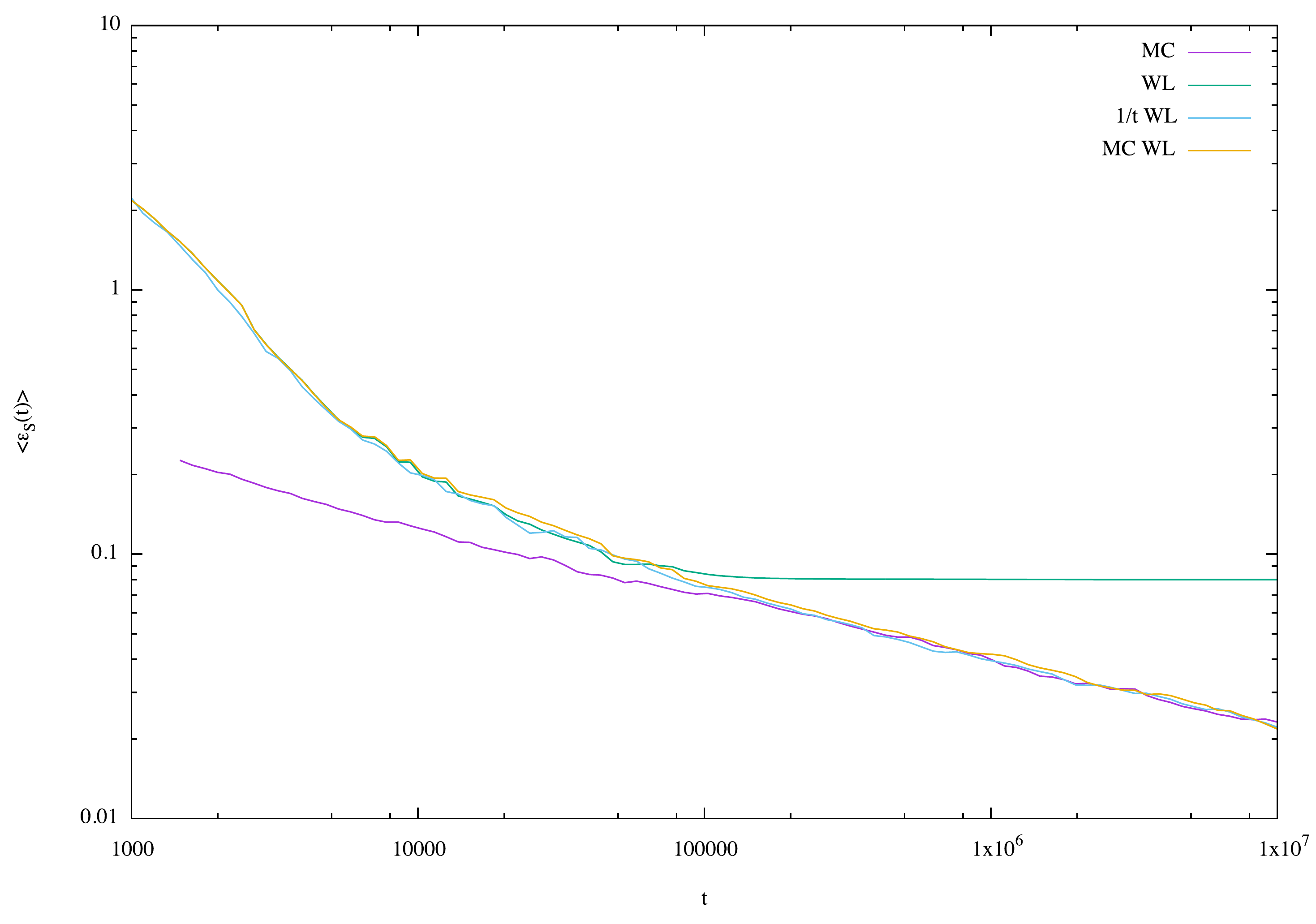}}
\centering\caption{Behavior of the errors in DOS ($S(E)=\ln g(E)$) as a function of simulation time $t$ for the $8\times8$ Ising model
using the MC (purple line), WL (green line), $1/t$ WL (blue line) and the new method labeled as "MC WL" (yellow line). Each curve is
the average of 100 independent runs. See text for more details.}
\label{fig1}
\end{figure}

\begin{figure}
\centering{\includegraphics[width=1.0\textwidth]{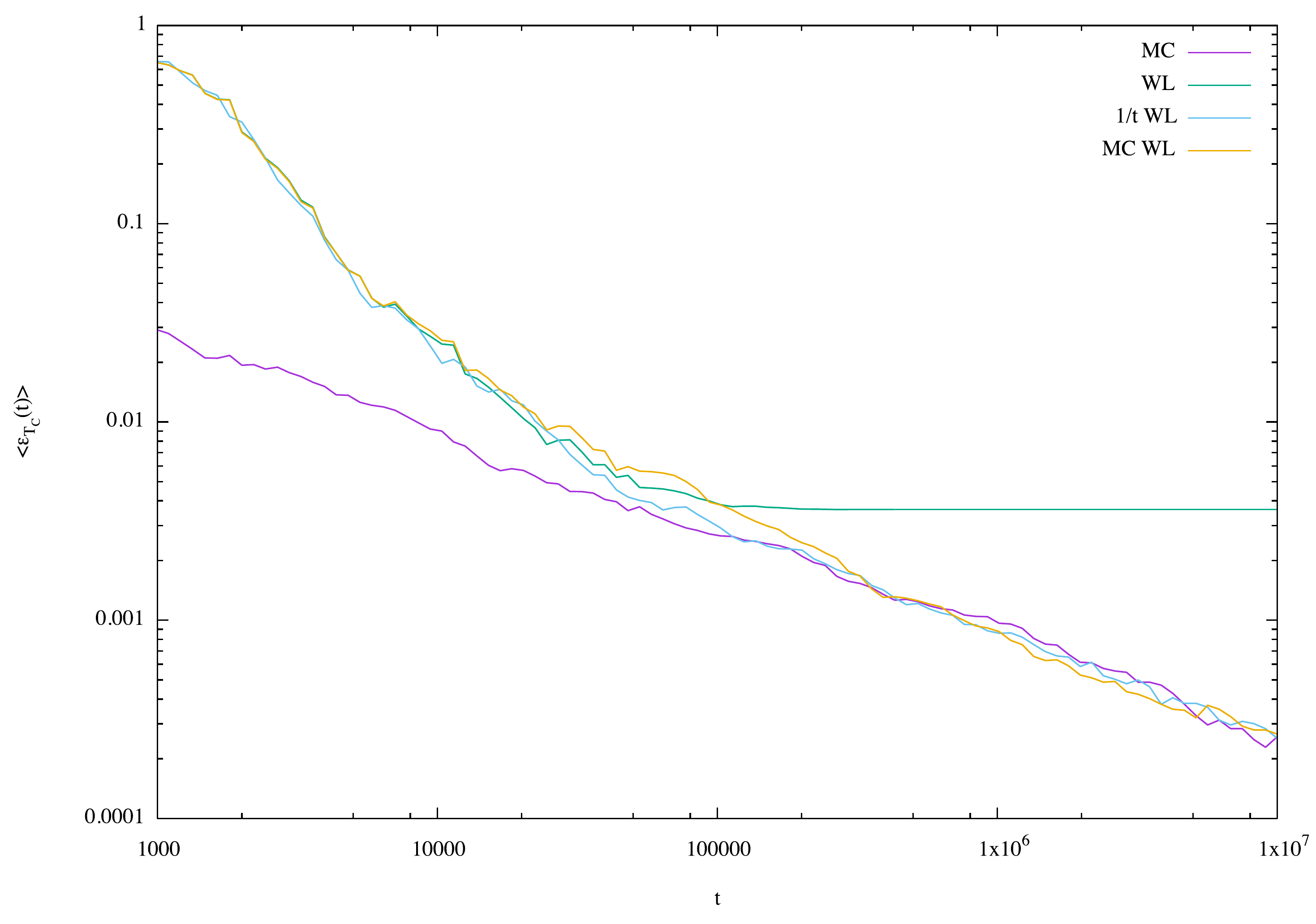}}
\centering\caption{Same as Fig. 1, but for the critical temperature ($T_C$). See text for more details.}
\label{fig1}
\end{figure}

\begin{figure}
\centering{\includegraphics[width=1.0\textwidth]{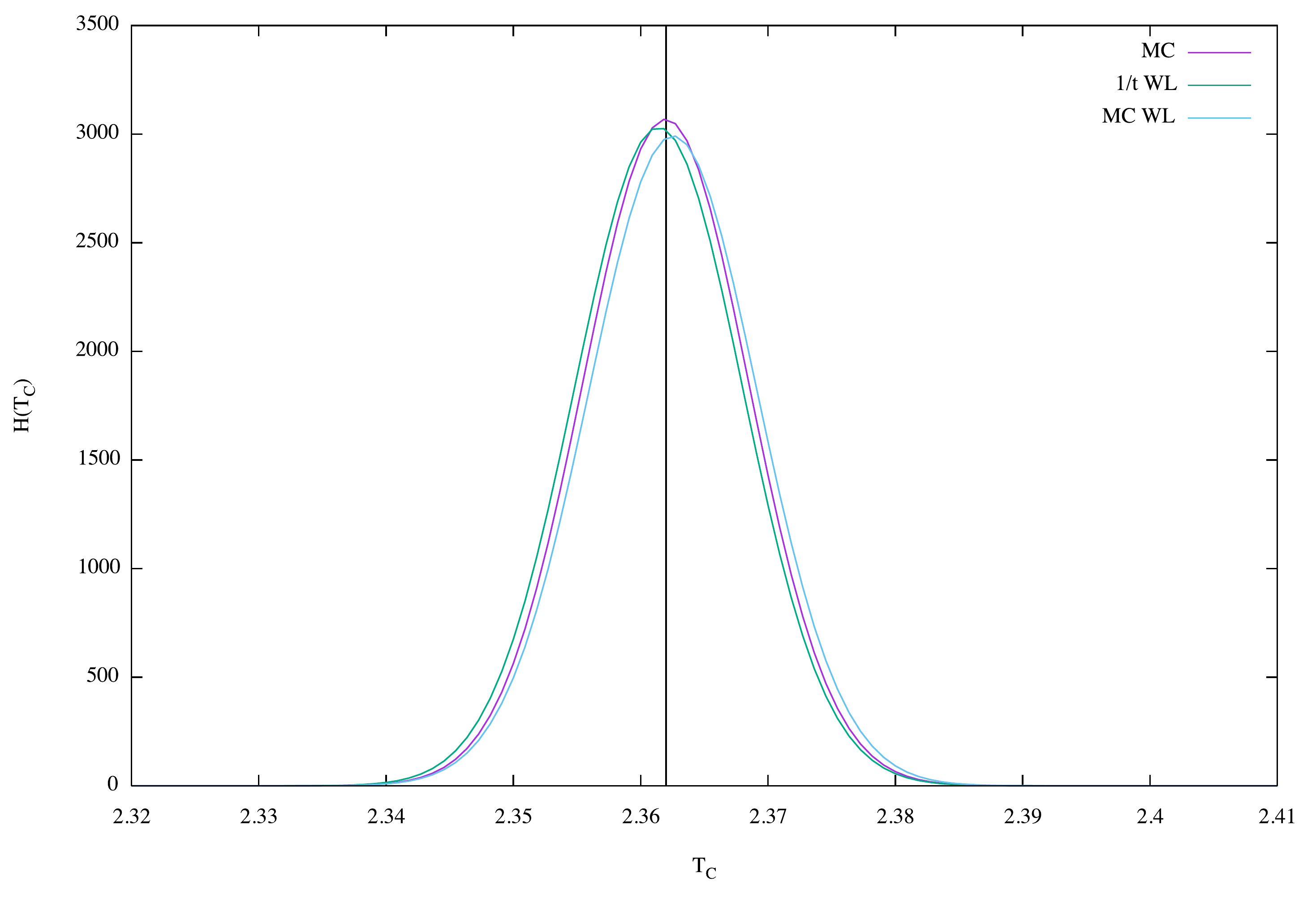}}
\centering\caption{Best-fit for the histograms of the critical temperature obtained using MC (purple line), $1/t$ WL (green line)
and MC WL (blue line) simulations. Each curve correspond to 50000 independent simulation runs of $10^7$ moves ($t=1.59\times10^6$).
The exact value ($T^{exc}_c=2.3620$) is marked as the vertical line.}
\label{fig1}
\end{figure}

\end{document}